\documentclass{ws-ijqi}
\usepackage{amssymb,amsmath}
\usepackage{mathrsfs}
\def\d{\operatorname{d}}\def\<{\langle}\def\>{\rangle}
\def\Tr{\operatorname{Tr}}
\def\openone{1\!\!1}

\def\Base{\set{B}}
\def\grp#1{{\mathbf #1}}
\def\set#1{{\sf
#1}}
\def\dim{\operatorname{dim}} 
\def\Span{\operatorname{Span}}
\def\Cmplx{\mathbb
C}

\def\spc#1{\mathcal{#1}}
\def\rep#1{{\sf #1}}
\def\Bnd#1{\mathcal{B}(#1)} 
\def\Bndd#1,#2{\mathcal{B}(#1,#2)}

\def\Proof{\medskip\par\noindent{\bf Proof. }}
\def\qed{$\,\blacksquare$\par}
\newtheorem{theo}{Theorem}
\newtheorem{prop}{Proposition}

\begin{document}

\markboth{G. Chiribella, G. M. D'Ariano, P. Perinotti, and M. F. Sacchi}
{Maximum likelihood method for estimating a group of physical transformations}

\catchline{}{}{}{}{}

\title{MAXIMUM LIKELIHOOD ESTIMATION FOR A GROUP OF PHYSICAL TRANSFORMATIONS}

\author{GIULIO CHIRIBELLA\footnote{chiribella@fisicavolta.unipv.it}}


\author{GIACOMO MAURO D'ARIANO\footnote{dariano@unipv.it}}

\address{QUIT, Dip. di Fisica ``A. Volta'', Universit\`a di Pavia, Via Bassi 6\\
Pavia, 27100, Italy}

\author{PAOLO PERINOTTI\footnote{perinotti@fisicavolta.unipv.it}}

\author{MASSIMILIANO F. SACCHI\footnote{msacchi@unipv.it}
}

\address{QUIT, INFM, Via Bassi 6\\
Pavia, 27100, Italy}

\maketitle

\begin{history}
\received{10 6 2005}
\end{history}

\begin{abstract}
  The maximum likelihood strategy to the estimation of group
  parameters allows to derive in a general fashion optimal
  measurements, optimal signal states, and their relations with other
  information theoretical quantities. These results provide a deep
  insight into the general structure underlying optimal quantum
  estimation strategies. The entanglement between representation
  spaces and multiplicity spaces of the group action appear to be the
  unique kind of entanglement which is really useful for the optimal
  estimation of group parameters.
\end{abstract}


\section{Introduction} 
Since the beginning of quantum estimation
theory\cite{Helstrom76,Holevo}, the research about measurements and
estimation strategies for the optimal detection of physical parameters
has been a major focus. In particular, the case where the physical
parameters to be estimated correspond to the unknown action of some
symmetry group has received a constant
attention\cite{Statistics,CloningAndStateEst,OptClocks}. The reason of
such interest, that in the last ten years received a strong motivation
from the new field of quantum information, is the broad spectrum of
applications of the topic, ranging from quantum statistics to quantum
cryptography, from the study of uncertainty relations to the design of
high sensitivity measurements achieving the ultimate quantum limit.

Despite the long-dated attention to the problem, many new examples
and applications of group parameter estimation came up recently, and,
on the other hand, some controversial points have been clarified only
in the very last years. In the rich variety of this scenario, it is
somehow natural to look for a general point of view, suitable for
capturing the main features of optimal estimation without entering the
specific details of the symmetry group involved in the particular
problem at hand. 

In this paper, we will show how such kind of general insight can be
provided in a simple way in the \emph{maximum likelihood
approach}\cite{MlPovms}, where the measurements are designed to
maximize the probability of estimating the true value of the unknown
group parameter. In this approach, the relations between the quality
of the estimation and other information theoretic properties, such as
the Holevo $\chi-$quantity\cite{HolevoChi} and the dimension of the
space spanned by a quantum state under the action of the group become
straightforward.

The results of the maximum likelihood method recently allowed to
understand the crucial role of the equivalent irreducible
representations of the group in the optimal estimation
strategy\cite{refframe}, giving a striking application of this
mechanism with the solution of a long-standing
controversy\cite{frames} about the efficiency in the absolute
transmission of a Cartesian reference frame. In that application, the
technique of equivalent representations has been the key idea for an
efficient use of quantum resources.

Here we will show in a general fashion that, far from being a
technicality, the use of equivalent representations is synonymous of
the use of a unique kind of entanglement that is really suitable for
group parameter estimation. More precisely, such kind of entanglement
is the entanglement between spaces where the group acts irreducibly
(\emph{representation spaces}) and spaces where the group acts
trivially (\emph{multiplicity spaces}). This entanglement is related
only to the group, without any relation with other ``natural'' tensor
product structures that can be present in the system, e.g. when the
system is made by a set of distinguishable particles.

The concept of representation/multiplicity is well known in the field
of quantum error correction\cite{Viola,Zanardi}, where the
multiplicity spaces are referred to as \emph{decoherence-free
subspaces} and, more generally, \emph{noiseless subsystems}. Moreover,
the same concept has recently found applications also in the context
of quantum communication\cite{QInfoWithoutRefframe} and
cryptography\cite{CrypthoRefframe}. On the other hand, the application
to quantum estimation of tools such as representation/multiplicity
spaces and entanglement is a completely new
issue\cite{refframe,HayashiSu2}.  In the maximum likelihood
approach the maximal entanglement between representation and
multiplicity spaces appears to be the unifying feature of the optimal
strategy for any group parameter estimation. Moreover, the validity of
this result is not limited to the maximum likelihood estimation, and
can be extended to estimation schemes where different figures of merit
are considered \cite{EntEstimation}.

In Section 2 we will present the general approach to group parameter
estimation. We will firstly start with a brief self-contained
introduction about group parameter estimation (2.1), maximum
likelihood approach (2.2), group theoretical tools (2.3), and
covariant measurements (2.3). After that, we will derive in the
general setting the optimal measurements (2.5), and the optimal states
(2.6), emphasizing the role of equivalent representations and the
relations with the Holevo $\chi-$quantity. Finally we will conclude
the section with a discussion (2.7) about the use of entanglement with
an additional reference system, showing how to efficiently use such
resource.

In Section 3, all the results will be generalized to infinite
dimension (Subsection 3.1) and non-compact groups (Subsection 3.2).
As an example of application, in Subsection 3.3 we provide a detailed
analysis of the estimation of the joint displacement of two harmonic
oscillators in the phase space, underlying in this continuous variable
context the connections with the well known example by Gisin and
Popescu about the estimation of a direction using parallel and
anti-parallel spins\cite{GisPop}.
\section{A general approach to group parameter estimation}
\subsection{Background}
The issue of this paper is the problem of optimally discriminating a
family of quantum signal states, which is the orbit generated by a
given input state under the action of a group. In other words, if the
input state is the density matrix $\rho \in \Bnd {\spc H}$ on the
Hilbert space $\spc H$, then we want to find the best estimation of
the states in the orbit
\begin{equation}\label{OrbitRho}
\mathcal{O}=\{~\rho_g= U_g \rho U_g^{\dag}~|~g \in \grp G\}~,
\end{equation} 
obtained by transforming the input state with the unitary
representation $\{U_g\}$ of the group $\grp G$.

In general, the points of the orbit are not in one-to-one
correspondence with the elements of the group, since one can have
$\rho_{g_1}=\rho_{g_2}$ even for different $g_1$ and $g_2$. However,
in this paper we will consider for simplicity the case where the
correspondence between the group and the orbit is one-to-one, since, a
part from a technical complication in the notation, the extension of
the results to the general case is straightforward.

In the case of one-to-one correspondence between signal states and
elements of the group, the problem of state estimation becomes
equivalent to estimating the action of a black-box that performs an
unknown unitary transformation drawn from the set $\{U_g~|~g \in \grp
G\}$. From this point of view it is also important to find the optimal
input states that allow to discriminate the action of the unitary
operators $\{U_g\}$ in the best possible way.

Whatever point of view we choose, we always need to estimate the value
of a group parameter. In order to do this, the most general estimation
strategy allowed by quantum mechanics is described by a Positive
Operator Valued Measure (POVM) $M$, that associates to any estimated
parameter ${\hat g} \in \grp G$ a positive semidefinite operator
$M({\hat g})$ on $\spc H$, satisfying the normalization condition
\begin{equation}\label{Norm}
\int_{\grp G}~ \d g~ M(g)= \openone~,
\end{equation} 
$\d g$ being the normalized invariant Haar measure on the group
($\int_{\grp G} \d g=1$\ \  $\d (hg)~=~\d (gh)~=~\d g$ \ \ $\forall g,h
\in \grp G$).  The probability density of estimating ${\hat g}$ when
the true value of the parameter is $g$ is then given by the usual Born
rule:
\begin{equation}\label{Born}
p({\hat g}|g)= \Tr[\rho_g ~M({\hat g})]~,
\end{equation}
where $\rho_g =U_g \rho U_g^{\dag}$.  Note that here we are
considering $\grp G$ as a continuous group only for fixing notation,
nevertheless---here and all throughout the paper---$\grp G$ can have a
finite number of elements, say $|\grp G|$, and in this case we have
simply to replace integrals with sums and $\d g $ with $1/|\grp G|$.

In order to find an optimal estimation strategy, we need firstly to
fix our optimality criterion. The most common way to do this is to
weigh the estimation errors with some {cost
function}\cite{Helstrom76,Holevo} $f(\hat g,g)$, that assesses the
cost of estimating $\hat g$ when the true value is $g$. Once the cost
function is fixed, we can adopt two different settings for the
optimization, the Bayesian and the frequentistic one. In the Bayesian
setting, one assumes a prior distribution of the true values (which is
usually uniform) and then the optimal estimation is the one that
minimizes the average cost, where the average is performed with
respect to the probability distributions of both the estimated and the
true values. On the other hand, in the frequentistic setting no prior
distribution of the true values is assumed, and one performs a
\emph{minimax} optimization, i.~e. minimizing the maximum (over all
possible true values) of the average cost, where now the average is
done just over the estimated values.

\subsection{Maximum likelihood approach}\label{MaxLikApproach}
Many different criteria can be used to define what is an optimal
estimation, each of them corresponding to a different choice of the
cost function in the optimization procedure. In general, the choice of
a cost function depends on the particular problem at hand. For
example, if we need to estimate a state, a natural cost is the
opposite of the fidelity between the estimated state and the true one,
while, if we are interested in the estimation of a parameter, a more
appropriate cost function would be the variance of the estimated
values.

In this paper, however, since we want to investigate general
properties of covariant estimation, we seek a criterion that maintains
a clear meaning in the largest number of different situations. The
simplest approach that allows a general analytical solution is given
by the \emph{maximum likelihood criterion}\cite{Helstrom76,Holevo},
which corresponds to the maximization of the probability (probability
density in the continuous case) that the estimated value of the
unknown parameter actually coincides with its true value.  In this
case, the cost function is a Dirac-delta $f(\hat g,g)=- \delta (\hat
g,g)$ (Kronecker-delta in the finite case). 

For finite groups maximum likelihood is in some sense the most natural
criterion. In fact, if we are trying to decide among a finite set of
alternatives which is the true one, of course we would like to do this
with the maximum probability of success. On the other hand, in the
continuous case, the maximum likelihood approach can still be
interpreted as the maximization of the probability that the estimated
value lies in a narrow neighborhood of the true one.

\subsection{Basic results from group theory}\label{BasicGrpTheo}
Here we will recall some useful tools from group theory that we will
exploit throughout the paper.

Consider a finite dimensional Hilbert space $\spc H$ and a unitary
(or, more generally, projective) representation $\rep R (\grp G)=
\{U_g\in \Bnd {\spc H}~|~g \in \grp G\}$ of a compact Lie group $\grp
G$.  The Hilbert space can be decomposed into orthogonal subspaces in
the following way
\begin{equation}\label{SpaceDecomp}
\spc H \equiv \bigoplus_{\mu \in \set S}~ \spc H_{\mu} \otimes \Cmplx^{m_{\mu}}~,  
\end{equation}
where the sum runs over the set $\set S$ of irreducible
representations of $\grp G$ that appear in the Clebsch-Gordan
decomposition of $\rep R(\grp G)$.  The action of the group is
irreducible in each \emph{representation space} $\spc H_{\mu}$, while
it is trivial in the \emph{multiplicity space} $\Cmplx^{m_{\mu}}$,
namely
\begin{equation}\label{RepDecomp}
U_g \equiv \bigoplus_{\mu \in \set S}~U_g^{\mu} \otimes \openone_{m_{\mu}}~,
\end{equation}
where each representation $\{U_g^{\mu}\}$ is irreducible, and
$\openone_{d}$ denotes the identity operator in a $d-$dimensional
Hilbert space.  Moreover, any operator $O \in \Bnd {\spc H}$ in the
commutant of $\rep R(\grp G)$---i.e. such that $[O,U_g]=0 \quad
\forall g \in \grp G$---has the form
\begin{equation}\label{CommutingOp}
O= \bigoplus_{\mu \in \set S}~ \openone_{d_{\mu}} \otimes O_{\mu}~,
\end{equation}
where $d_{\mu}$ is the dimension of $\spc H_{\mu}$, and $O_{\mu}$ is a
$m_{\mu} \times m_{\mu}$ complex matrix.  In particular, the group
average $\<A\>_{\grp G}\equiv \int \d g~ U_g A U_g^{\dag}$ of a given
operator $A$ with respect to the invariant Haar measure $\d g$ is in
the commutant of $\rep R(\grp G)$, and has the form:
\begin{equation}\label{AveOp}
\<A\>_{\grp G}=\bigoplus_{\mu \in \set S}~ \openone_{d_{\mu}} \otimes
\frac{1}{d_{\mu}}~ \Tr_{\spc H_{\mu}}[A]~,
\end{equation}
where $\Tr_{\spc H_{\mu}}[A]$ is a short notation for $\Tr_{\spc
H_{\mu}}[P_\mu AP_\mu]$, $P_\mu$ denoting the orthogonal projector
over the Hilbert subspace $\spc H_{\mu} \otimes \Cmplx^{m_{\mu}}$ in
the decomposition (\ref{SpaceDecomp}) of $\spc H$.  Here and
throughout the paper we assume the normalization of the Haar measure:
$\int_{\grp G} \d g=1$.

\bigskip
{\bf Remark I:} \emph{entanglement between representation spaces and
multiplicity spaces.}\\ The choice of an orthonormal basis
$\Base^{\mu}~=~\{|\phi_n^{\mu}\>~\in~\Cmplx^{d_{\mu}}~|~n=1, \dots
,m_{\mu}\}$ for a multiplicity space fixes a particular decomposition
of the Hilbert space as a direct sum of irreducible subspaces:
\begin{equation}\label{DirectSum}
\spc H_{\mu} \otimes \Cmplx^{m_{\mu}} = \oplus_{n=1}^{m_{\mu}}~ \spc H^{\mu}_n~,
\end{equation}
where $\spc H_n^{\mu} \equiv \spc H_{\mu} \otimes |\phi_n^{\mu}\>$.
In this picture, it is clear that $m_\mu$ is the number of different
irreducible subspaces carrying the same representation $\mu$, each of
them having dimension $d_{\mu}$.  Moreover, with respect to the
decomposition (\ref{SpaceDecomp}), any pure state $|\Psi\> \in \spc H$
can be written as
\begin{equation}\label{StateDecomp}
|\Psi\>= \bigoplus_{\mu \in \set S}~ c_{\mu}~|\Psi_{\mu}\>\!\>~,
\end{equation} 
where $|\Psi_{\mu}\>\!\>$ is a bipartite state in $\spc H_{\mu}
\otimes \Cmplx^{m_{\mu}}$ and $\sum_{\mu \in \set
S}~|c_{\mu}|^2=1$. With respect to the direct sum decomposition
(\ref{DirectSum}), the Schmidt number of such a state is the minimum
number of subspaces carrying the same representation $\mu \in \set S$
that are needed to decompose $|\Psi\>$.
 
\bigskip
{\bf Remark II:} \emph{maximum number of equivalent representations in
the decomposition of a pure state.}\\ The Schmidt number of any
bipartite state $|\Psi_{\mu}\>\!\>~\in~\spc
H_{\mu}~\otimes~\Cmplx^{m_{\mu}}$ is always less then or equal to
$k_{\mu}= \min \{d_{\mu}, m_{\mu}\}$.  This means that any pure state
can be decomposed using \emph{no more} than $k_{\mu}$ irreducible
subspaces carrying the same representation $\mu \in \set S$.

\subsection{Covariant measurements}
Since the set of states to be estimated is invariant under the action
of the group, there is no loss of generality in assuming a covariant
POVM, i.e. a POVM satisfying the property $M(hg)=U_h M(g) U_h^{\dag}$
for any $g,h \in \grp G$. In fact, it is well known that, for any
possible POVM, there is always a covariant one with the same average
cost\cite{Holevo}, this result holding both in the minimax approach
and in the Bayesian approach with uniform prior distribution.

A covariant POVM has the form
\begin{equation}\label{CovPovm}
M(g)=U_{g} \Xi U_{g}^{\dag}~,
\end{equation} 
where $\Xi$ is a positive semidefinite operator.  For covariant
POVM's, exploiting the formula (\ref{AveOp}) for the group average,
the normalization condition (\ref{Norm}) can be translated into a
simple set of conditions for the operator $\Xi$:
\begin{equation}\label{TrXi}
\Tr_{\spc H_{\mu}}[\Xi]=d_{\mu}~ \openone_{m_{\mu}}~.
\end{equation}
In this way, the optimization of a covariant POVM is reduced to the
optimization of a positive operator satisfying the constraints
(\ref{TrXi}).

\subsection{Optimal measurements}\label{OptMeasurements}
Here we derive for any given input state $|\Psi\>\in \spc H$ the
measurement that maximizes the probability (density) of estimating the
true value of the unknown group parameter $g~\in~\grp G$. Note that,
due to covariance, this probability has the same value for any group
element: $p(g|g)=\<\Psi|\Xi|\Psi\>$, according to Eqs. (\ref{Born})
and (\ref{CovPovm}). In order to find the POVM, it is convenient to
express the input state in the form (\ref{StateDecomp}), and write
each bipartite state $|\Psi_{\mu}\>\!\>$ in the Schmidt form:
\begin{equation}\label{Schmidt}
|\Psi_{\mu}\>\!\>= \sum_{m=1}^{r_{\mu}}~\sqrt{\lambda^{\mu}_m}~
 |\psi_m^{\mu}\>|\phi_m^{\mu}\>~,
\end{equation}  
where $r_\mu\le k_\mu=\min \{d_{\mu},m_{\mu}\}$ is the Schmidt number, and $\lambda_m^{\mu}>0 \quad \forall \mu ,m$.  We can now define
the projection 
\begin{equation}\label{P_Psi}
P_{\Psi}=\bigoplus_{\mu \in \set
S}~\sum_{m=1}^{r_{\mu}}~\openone_{d_{\mu}} \otimes
|\phi_m^{\mu}\>\<\phi_m^{\mu}|~.
\end{equation}
It projects onto the subspace $\spc H_{\Psi}$ spanned by the orbit of
the input state, this subspace being also the smallest invariant
subspace containing the input state.

Clearly, the probability distribution of the outcomes of a covariant
measurement $M(\hat g)=U_{\hat g}~\Xi~U_{\hat g}^{\dag}$ performed on
any state in the orbit depends only on the projection
$P_{\Psi}~\Xi~P_{\Psi}$. Therefore, to specify an optimal covariant
POVM for the state $|\Psi\>$, we need only to specify the operator
$P_{\Psi}\Xi P_{\Psi}$. All covariant POVM's corresponding to the same
operator will be equally optimal.  \begin{theo}[optimal
POVM]\label{TheoOptPOVM} For a pure input state $|\Psi\>$, the optimal
covariant POVM in the maximum likelihood approach is given by
\begin{equation}\label{OptXi}
P_{\Psi} \Xi P_{\Psi}= |\eta\>\<\eta|~,
\end{equation}
where 
\begin{equation}\label{OptEta}
|\eta\>= \bigoplus_{\mu \in \set S}~ \sqrt{d_{\mu}}e^{i \arg
 (c_{\mu})}~ \sum_{m=1}^{r_{\mu}} |\psi_m^{\mu}\>|\phi_m^{\mu}\>~.
\end{equation}
The value of the likelihood for the optimal POVM is
\begin{equation}\label{OptLik}
p^\mathrm{Opt}(g|g)= \left( \sum_{\mu \in \set S}
|c_{\mu}|\sum_{m=1}^{r_{\mu}} \sqrt{\lambda_m^{\mu} d_{\mu}} \right)^2
\qquad \forall g~.
\end{equation}
\end{theo}
\Proof Using Schwartz inequality, the likelihood can be bounded as
follows:
\begin{equation*}
\begin{split}
p(g|g)&=\<\Psi|\Xi|\Psi\>\\
&\leq \sum_{\mu,\nu}~| c_{\mu}c_{\nu}|~\left| \<\!\<\Psi_{\mu}| \Xi |\Psi_{\nu}\>\!\> \right|\\
&\le \left ( \sum_{\mu} |c_{\mu}|~\sqrt{\<\!\<\Psi_{\mu}|\Xi|\Psi_{\mu}\>\!\>} \right)^2 ~.
\end{split}
\end{equation*}
Moreover, exploiting the Schmidt form (\ref{Schmidt}) and applying a
second Schwartz inequality, we obtain
\begin{equation*}
\begin{split}
\<\!\<\Psi_{\mu}|\Xi |\Psi_{\mu}\>\!\>&=\sum_{m,n=1}^{r_{\mu}}~
\sqrt{\lambda_m^{\mu} \lambda_n^{\mu}}
~\<\psi_m^{\mu}|\<\phi_m^{\mu}|~\Xi~|\psi_n^{\mu}\>|\phi_n^{\mu}\>\\
&\leq \left( \sum_{m=1}^{r_{\mu}}~\sqrt{\lambda_m^{\mu}~~
\<\psi_m^{\mu}|\<\phi_m^{\mu}|~\Xi~|\psi_m^{\mu}\>|\phi_m^{\mu}\>}\right)^2~.
\end{split}\end{equation*}
Finally, the positivity of $\Xi$ implies 
\begin{equation*}
\begin{split}
\<\psi_m^{\mu}|\<\phi_m^{\mu}|~\Xi~|\psi_m^{\mu}\>|\phi_m^{\mu}\> &\le
\<\phi_m^{\mu}| ~\Tr_{\spc H_{\mu}} [\Xi]~|\phi_m^{\mu}\>=d_{\mu}~,
\end{split}\end{equation*}
 due to the normalization condition (\ref{TrXi}).  By putting together
 these inequalities, we obtain the bound
\begin{equation*}p(g|g) \le \left( \sum_{\mu \in \set S} |c_{\mu}| \sum_{m=1}^{r_{\mu}}~ \sqrt{\lambda_m^{\mu} d_{\mu}} \right)^2\equiv p^\mathrm{Opt}(g|g)~,
\end{equation*} holding for any possible POVM. It is immediate to see that the covariant POVM given by (\ref{OptXi},\ref{OptEta}) achieves the bound, hence it is optimal. \qed  

\bigskip
{\bf Remark I:} \emph{uniqueness of the optimal POVM.}\\ Since the
Theorem specifies the optimal POVM only in the subspace spanned by the
orbit of the input space, it follows that the optimal POVM is unique
if and only if the orbit spans the whole Hilbert space.  If it is not
the case, one can arbitrarily complete the POVM given by (\ref{OptXi})
to the entire Hilbert space.

\bigskip
{\bf Remark II:} \emph{square-root measurements.}\\ The optimal POVM
in Eqs. (\ref{OptXi}) and (\ref{OptEta}) coincides with the so-called
``square-root measurement''\cite{goodpovm}. In fact, such a
measurement has a POVM with $|\eta\>=F^{-\frac{1}{2}}|\psi\>$, with
the ``frame operator'' $F$ given by $F=\int\d g
U_g|\psi\>\<\psi|U_g^\dag$. Using Eq. (\ref{AveOp}) one finds
$F=\bigoplus_\mu |c_\mu|^2\sum_m \frac{\lambda_m^\mu}{d_\mu}
\openone_\mu\otimes |\phi_m^\mu\>\<\phi_m^\mu|$, and one can easily
check that $|\eta\>=F^{-\frac{1}{2}}|\psi\>$.

\subsection{Optimal input states}\label{OptInputStates}
While in the previous paragraph we assumed the input state to be
given, and we were mainly interested in the problem of state
estimation, here we will focus our attention on the problem of
estimating the action of a black box that performs an unknown unitary
transformation drawn from a group.  From this point of view, our aim
is now to determine which are the states in the Hilbert space that
allow to maximize the probability of successfully discriminating the
unknown unitaries $\{U_g\}$.

We will first show that the dimension of the subspace spanned by the
orbit of the input state is always an upper bound for the likelihood,
and that this bound can always be achieved by using suitable input
states. Then, the optimal input states will be the ones that maximize
the dimension of the subspace spanned by the orbit.
\begin{lemma}
Let $d_{\Psi}=\dim \Span \{U_g|\Psi\>~|~g \in \grp G\}$ be the
dimension of the subspace spanned by the orbit of the input
state. Then
\begin{equation}\label{Dimension}
d_{\Psi}=\sum_{\mu \in \set S}  d_{\mu} r_{\mu}~,
\end{equation}
where $r_{\mu}$ is the Schmidt number of the bipartite state $|\Psi_{\mu}\>\!\>$ (we define $r_{\mu}=0$ if $|c_{\mu}|=0$ in the decomposition (\ref{StateDecomp})~).
\end{lemma} 
\Proof The subspace spanned by the orbit is the support of the frame
operator
\begin{equation*}
 \int_{\grp G} \d g~ U_g|\Psi\>\<\Psi|U_g^{\dag}= \bigoplus_{\mu \in \set S} |c_{\mu}|^2 \openone_{d_{\mu}} \otimes \frac{\Tr_{\spc H_{\mu}} [|\Psi_{\mu}\>\!\>\<\!\<\Psi_{\mu}|]}{d_{\mu}}~,
\end{equation*}
the r.h.s. coming from Eq. (\ref{AveOp}).  Using the Schmidt form
(\ref{Schmidt}) of each bipartite state $|\Psi_{\mu}\>\!\>$, it follows that the dimension of the support is $d_{\Psi}=\sum_{\mu \in \set
S} d_{\mu} r_{\mu}$. \qed
\begin{theo}[relation between likelihood and dimension]\label{TheoBoundDimension}
For any pure input state $|\Psi\> \in \spc H$, the following bound holds:
\begin{equation}\label{BoundDimension}
p(g|g) \le d_{\Psi}~.
\end{equation}
The bound is achieved if and only if the state has the form
\begin{equation}\label{OptStateR}
|\Psi\>=\frac{1}{\sqrt{d_{\Psi}}}~\bigoplus_{\mu \in \set S}~ \sqrt{d_{\mu} r_{\mu}}~e^{i \theta_{\mu}}~ |\Psi_{\mu}\>\!\>~,
\end{equation}
where $e^{i\theta_{\mu}}$ are arbitrary phase factors and
$|\Psi_{\mu}\>\!\>~\in~\spc H_{\mu}~\otimes~\Cmplx^{m_{\mu}}$ is a
bipartite state with Schmidt number $r_{\mu}$ and equal Schmidt
coefficients ($\lambda_{m}^{\mu}=1/r_{\mu}$ for any $m=1, \dots,
r_{\mu}$).
\end{theo}

\Proof
Exploiting Eq.(\ref{OptLik}), we have 
\begin{eqnarray}
\label{First}p(g|g)&\le& p^\mathrm{Opt}(g|g)\\
&=&\left( \sum_{\mu} ~|c_{\mu}|~ \sum_{m=1}^{r_{\mu}} \sqrt{\lambda_m^{\mu} d_{\mu}}\right)^2\\
\label{Second}&\le& \left( \sum_{\mu} ~|c_{\mu}|~\sqrt{r_{\mu} d_{\mu}} \right)^2\\
\label{Fourth}&\le& \sum_{\mu} r_{\mu} d_{\mu} = d_{\Psi}~,
\end{eqnarray}
the inequalities (\ref{Second}) and (\ref{Fourth}) coming from
Schwartz inequality and from the normalizations $\sum_{m=1}^{r_{\mu}}
\lambda_m^{\mu}=1$ and $\sum_{\mu} |c_{\mu}|^2=1$.  Let us see when
this bound is attained. Clearly, the equality in (\ref{First}) holds
if we use the optimal POVM of Theorem \ref{TheoOptPOVM}. On the other
hand, the Schwartz inequality (\ref{Second}) becomes equality if and
only if $\lambda_m^{\mu}= 1/r_{\mu}$ for any $m=1, \dots,
r_{\mu}$. Finally, the last Schwartz inequality (\ref{Fourth}) becomes
equality if and only if
$|c_{\mu}|=\sqrt{\frac{r_{\mu}d_{\mu}}{d_{\Psi}}}$. The requirements
$|c_{\mu}|=\sqrt{ \frac{ r_{\mu} d_{\mu} }{d_{\Psi}}}$ and
$\lambda_m^{\mu}= 1/r_{\mu}$ are satisfied only by states of the form
(\ref{OptStateR}). \qed We can now answer to the question which are
the best input states for discriminating a group of unitaries.
\begin{theo}[optimal input states]\label{TheoOptState}
For any state $\rho$ on $\spc H$ and for any POVM, the likelihood is
bounded from above by the quantity
\begin{equation}\label{OptStateLik}
L= \sum_{\mu \in S}~ d_{\mu} k_{\mu}~,
\end{equation}
where $k_{\mu} \equiv \min \{d_{\mu},m_{\mu}\}$.
The bound is achieved by pure states of the form
\begin{equation}\label{OptState}
|\Psi\>=\frac{1}{\sqrt{L}}~\bigoplus_{\mu \in \set S}~ \sqrt{d_{\mu} k_{\mu}}~e^{i \theta_{\mu}}~ |E_{\mu}\>\!\>~,
\end{equation}
where $e^{i\theta_{\mu}}$ are arbitrary phase factors and $|E_{\mu}\>\!\>~\in~\spc H_{\mu}~\otimes~\Cmplx^{m_{\mu}}$ are arbitrary maximally entangled states.
\end{theo}
\Proof Since the likelihood $\mathcal L [\rho]=\Tr[\rho \Xi]$ is a
linear functional of the input state, it is clear that the maximum
likelihood over all possible states is achieved by a pure state.
Therefore, according to Eq. (\ref{BoundDimension}), the maximum
likelihood is given by the maximum of $d_{\Psi}$ over all pure
states. Since the Schmidt number $r_{\mu}$ cannot exceed
$k_{\mu}~=~\max \{d_{\mu},m_{\mu}\}$, we obtain that the maximum value
is

\begin{equation*}L~=~\max\{d_{\Psi}~|\quad|\Psi\> \in \spc H\}~=~\sum_{\mu \in \set S}~d_{\mu} k_{\mu}~.
\end{equation*} According to Theorem
(\ref{TheoBoundDimension}), such a maximum is achieved by pure states
of the form (\ref{OptState}).\qed

The results of Theorems \ref{TheoOptPOVM}, \ref{TheoBoundDimension},
and \ref{TheoOptState} have some important consequences.

{\bf Consequence I (each irreducible subspace contributes to the
likelihood with its dimension)}\\ According to Eq. (\ref{OptState}),
the probability of successful discrimination is maximized by
exploiting in the input state \emph{all the irreducible
representations} appearing in the Clebsch-Gordan decomposition of
$U_g$. Moreover, the contribution of each irreducible subspace to the
likelihood is related to the dimension $d_{\mu}$ by
Eqs. (\ref{Dimension}), (\ref{BoundDimension}), and
(\ref{OptStateLik}). In other words, the maximum likelihood approach
allows to give a general quantitative formulation to the common
heuristic argument that relates the quality of the estimation to the
dimension of the subspace spanned by the orbit of the input
state. From this point of view, the interpretation of the well known
example\cite{GisPop} about the quantum information of two parallel vs
anti-parallel spin $1/2$ particles is clear: for parallel spins the
input state lies completely in the triplet (symmetric) subspace, while
for anti-parallel it has a nonzero component also onto the
singlet. Evaluating the likelihood with Eq. (\ref{OptLik}), we have
indeed $p(g|g)^\mathrm{Opt}= (1 +\sqrt{3})^2/2 \approx 3.73$ for
anti-parallel spins, instead of $p(g|g)^\mathrm{Opt}=3$ for parallel
ones.  Notice, however, that the latter is not the optimal input state
in the maximum likelihood approach, which instead has coefficients
$c_\mu=\sqrt{\frac{r_\mu d_\mu}{d_{\Psi}}}$, with $\mu=0$ denoting the
singlet and $\mu=1$ the triplet. The largest $r_\mu$ are given by
$r_\mu=1$, whence $c_\mu= \sqrt{\frac{d_\mu}{d_{\Psi}}}$, namely
$c_0=\frac{1}{2}$ and $c_1=\frac{\sqrt{3}}{2}$, giving likelihood
$p(g|g)^\mathrm{Opt}=4$. It is possible to show\cite{MlPovms} that
this optimal input state can be chosen as a factorized state, such a
state being the tensor product of two {\em mutually unbiased} spin
states.

{\bf Consequence II (key role of equivalent representations)}\\ The
repeated use of equivalent representations is crucial for attaining
the maximum probability of successful discrimination. In fact, in
order to achieve the upper bound (\ref{OptStateLik}) one necessarily
needs to use the maximal amount of entanglement between representation
spaces and multiplicity spaces, corresponding to the maximum number of
irreducible subspaces carrying the same representation $\mu$, for any
$\mu$ in the Clebsch-Gordan decomposition.

{\bf Consequence III (maximization of the Holevo $\chi$-quantity)}\\
The optimal states in the maximum likelihood approach are those which
maximize the Holevo $\chi$-quantity\cite{HolevoChi}, which in the
group covariant case is defined as
\begin{equation}\label{ChiDef}
\chi_{\grp G}(\rho)=S\left( \int \d g~ U_g \rho U_g^{\dag}\right)-
\int \d g~ S(U_g \rho U_g^{\dag})~,
\end{equation}
$S(\rho)=-\Tr[\rho \log (\rho)]$ being the von Neumann entropy.  In
fact, for pure input states $\rho=|\Psi\>\<\Psi|$, the $\chi$-quantity
is the entropy of the average state: $\chi_{\grp G}(\rho)=
S\left(\<~\rho~\>_{\grp G}\right)$. Using Eq. (\ref{AveOp}), we have
\begin{equation*}
\<~\rho~\>_{\grp G}=\bigoplus_{\mu \in \set S}~|c_{\mu}|^2
\frac{\openone_{d_{\mu}}}{d_{\mu}} \otimes \Tr_{\spc H_{\mu}}\left[
|\Psi_{\mu}\>\!\>\<\!\<\Psi_{\mu}| \right]~.
\end{equation*}
It is then easy to see that, for any pure state $\rho=|\Psi\>\<\Psi|$,
\begin{equation}
\chi_{\grp G}(\rho)\le \log d_{\Psi}~,
\end{equation}  and that the bound is attained by states of the form
(\ref{OptStateR}). Finally, the maximum over all pure states is
\begin{equation}
\chi_{\grp G}(\rho)= \log L~,
\end{equation}
achieved by states of the form (\ref{OptState}). In this way,
the likelihood is directly related to the $\chi$-quantity, providing an upper bound to the amount of classical information that
can be extracted from the orbit of the input state.
\section{Internal vs external entanglement}
Up to now we looked for the optimal input states in the Hilbert space
of the system undergoing the unknown group transformation. From this
point of view, the entanglement between representation and
multiplicity spaces was just a kind of \emph{internal entanglement},
between two \emph{virtual subsystems}\cite{Zanardi} with Hilbert
spaces given by the representation and the multiplicity spaces,
respectively.

Suppose now that we can exploit an additional entangled resource,
i.e. we can entangle the system that undergoes the unknown group
transformation with an additional external system, which acts as a
reference.  In this case, we have to consider the tensor product
Hilbert space $\spc H \otimes \spc H_{\mathcal{R}}$, where the group
acts via the representation $\{U_g'=U_g \otimes
\openone_{\mathcal{R}}~|~g \in \grp G\}$. From the point of view of
the group structure, the only effect of the reference system is simply
to increase the multiplicity of the irreducible representations. In
fact, the Clebsch-Gordan decompositions of $\{U_g\}$ and $\{U_g'\}$
contain exactly the same irreducible representations, while the new
multiplicities are $m_{\mu}'=m_{\mu} \cdot d_{\mathcal{R}}$, where
$d_{\mathcal{R}}=\dim \spc H_{\mathcal{R}}$.  Therefore, from the sole
consideration of the decomposition of the Hilbert space, we obtain the
following:
\begin{theo}[use of external entanglement] 
Once $m_{\mu}'=m_{\mu} \cdot d_{\mathcal{R}} \ge d_{\mu}$ for any
irreducible representation $\mu \in \set S$, any further increase of
the dimension of the reference system is useless.\\ In particular, if
$d_{\mu} \ge m_{\mu} \quad \forall \mu \in \set S$, there is no need
of entanglement with an external system.
\end{theo}
\Proof In the decomposition (\ref{StateDecomp}) of a pure state, the
rank of any bipartite state cannot exceed
$k_{\mu}'=\min\{d_{\mu},m_{\mu}'\}$. Therefore, once $m_{\mu}'\ge
d_{\mu}$ for any $\mu$, the orbit of any pure state can always be
embedded in a subspace where all the multiplicities are equal to the
dimensions.\qed In other words, once the saturation $m_{\mu}' \ge
d_{\mu}$ is reached, there is no need of increasing the dimension of
the reference system.
\begin{corollary}[dimension of the reference system]\label{CorExtEnt} The maximum dimension of an external system that is useful for estimation is
\begin{equation}
\bar d_{\mathcal{R}}= \max \left\{ \left\lceil
\frac{d_{\mu}}{m_{\mu}}\right\rceil~|~ \mu \in \set S \right\}~,
\end{equation}
where the ``ceiling'' $\lceil x \rceil$ denotes the minimum integer
greater than $x$.
\end{corollary}
This mechanism of \emph{saturation of the multiplicities} can be
simply quantified in terms of the likelihood. In fact, the improvement
coming from the external reference system can be evaluated using
Theorem \ref{TheoOptState}, yielding the value of the likelihood for
the optimal input state: $L'=\sum_{\mu \in \set S} d_{\mu} k_{\mu}'$,
where $k_{\mu}'=\min \{d_{\mu},m_{\mu}'\}$.  The upper bound that can
be achieved with the use of a reference system as in Coollary
\ref{CorExtEnt} is then
\begin{equation}
L_{\max}= \sum_{\mu \in \set S} d_{\mu}^2~.
\end{equation}

\section{Generalization to infinite dimension and non-compact groups}
\subsection{Compact groups in infinite dimension} 
The main problem with infinite dimension comes from the fact that, in
some cases, the optimal states of section \ref{OptInputStates} are not
normalizable. From a physical point of view, this means that one has
to approximate them with normalized states in some reasonable way,
fixing additional constraints as, for example, the energy
constraint. Clearly the best approximation depends on the particular
problem at hand.

In a similar way, the POVM elements $P(g)=U_g ~\Xi~U_g^{\dag}$ in
general are not operators. For example, the well known optimal POVM
for the estimation of the phase of the radiation field is given by
\begin{equation}
P(\phi)= |e(\phi)\>\<e(\phi)|~,
\end{equation} 
where $|e(\phi)\>=\sum_{n=0}^{\infty}~e^{i n \phi}~ |n\>$ are the
so-called Susskind-Glogower vectors\cite{SussGlo}. Since such vectors
are not normalizable, the POVM elements $P(\phi)$ are not operators
acting in the Hilbert space $\spc H$.  For this reason, in infinite
dimension one should substitute the positive operator $P(g)$ with a
positive {\em form} $\pi_g$, defined by $\pi_g
~(|\phi\>,|\psi\>)~=~\<\phi|P(g)|\psi\>$.

However, except for this technicality, all results of Section
\ref{OptMeasurements} concerning optimal POVM's are essentially valid
in infinite dimension.

\subsection{Non-compact groups} 
The generalization of our method to the case of non-compact groups is
more involved than for compact groups in infinite
dimension. Nevertheless, such a generalization is crucial for many
physically meaningful cases, e.~g. the estimation of displacement or
of squeezing parameters in quantum optics.

In the following, we will consider the case of \emph{unimodular
groups}, i.e. groups for which the left-invariant measure $\d_L g$
($\d_L h g=\d_L g~~\forall g,h \in \grp G$) and the right-invariant
one $\d_R g$ ($\d_R gh=\d_R g~~\forall g,h \in \grp G$) coincide.  We
will then define the invariant Haar measure as $\d g=\d_L g =\d_R g$.
Notice that, in the Bayesian point of view, it is no longer possible
to assume the group parameters to be distributed according to such a
measure, since the uniform measure over a non-compact group is
non-normalizable.

In general, for non-compact groups the irreducible representations
contained in the Clebsch-Gordan decomposition may form a continuous
set.  To deal with such a situation one should replace in the
Sec. \ref{OptMeasurements} and \ref{OptInputStates} direct sums with
direct integrals. For example, the decomposition of the Hilbert space
(\ref{SpaceDecomp}) would rewrite\cite{delirio}
\begin{equation}
\int_{\set S}^{\oplus} m(\d \mu)~ \spc H_{\mu} \otimes \spc M_{\mu}~,
\end{equation}
$m(\d \mu)$ being a measure over the set of irreducible
representations, and $\spc M_{\mu}$ denoting the multiplicity
space. In the following, we will not carry on this rather technical
generalization, leaving the case of a direct integral of irreducible
representations only to a specific example (see next paragraph). We
will instead consider the simplest case of group representations that
can be decomposed in a \emph{discrete series of irreducible
components}. In other words, we will assume that it is still possible
to write the Clebsch-Gordan decomposition
\begin{equation}\label{RepDecompInfty}
U_g = \bigoplus_{\mu \in \set S}~ U_g^{\mu} \otimes \openone_{\spc M_{\mu}}~,
\end{equation} where the set $\set S$ is discrete.
 
Finally, we require any irreducible representation
$\{U_g^{\mu}\}$ in the Clebsch-Gordan series to be \emph{square summable}, that, in the in case of unimodular groups, is equivalent to the property 
\begin{equation}\label{SquareSummable}
\int \d g~ \left| \<\psi_{\mu}|~U_g~|\phi_{\mu}\> \right|^2< \infty
\qquad \forall ~|\psi_{\mu}\>,|\phi_{\mu}\> \in \spc H_{\mu}~.
\end{equation}
Under these hypotheses, the results of Theorems \ref{TheoOptPOVM},
\ref{TheoBoundDimension}, and \ref{TheoOptState} can be immediately
extended to non-compact groups. In fact, in this case we can exploit a
simple generalization of the formula (\ref{AveOp}) for the group
average, which allows us the use all the results of of Sections
\ref{OptMeasurements} and \ref{OptInputStates} by just substituting
the dimensions $d_{\mu}$ of the irreducible subspaces with their
\emph{formal dimensions}. \begin{prop} Let be $\{U_g\}$ a discrete
series of square-summable representations of a unimodular group. Then
the group average $\<A\>_{\grp G}$ of a given operator $A$ is given by
\begin{equation}
\<A\>_{\grp G}= \bigoplus_{\mu \in \set S}~ \openone_{\spc H_{\mu}} \otimes \frac{\Tr_{\spc H_{\mu}} [A]}{d_{\mu}}~,
\end{equation} where the \emph{formal dimension} $d_{\mu}$ is defined as
\begin{equation}\label{FormDim}
d_{\mu}= \left( \int \d g ~ |\<\psi_{\mu}|~U_g^{\mu}~|\phi_{\mu}\>|^2
\right)^{-1}~,
\end{equation}
$|\psi_{\mu}\>$ and $|\phi_{\mu}\>$ being any two normalized states in
$\spc H_{\mu}$.
\end{prop}\Proof Since the
group average $\<A\>_{\grp G}$ of an operator is in the commutant of
the representation (\ref{RepDecompInfty}) it has the form $\<A\>_{\grp
G}= \bigoplus_{\mu}~ \openone_{\spc H_{\mu}} \otimes A_{\mu}$, for
some suitable operators $A_{\mu}$ acting in the multiplicity
space. Taking the expectation value with respect to a normalized
vector $|\psi_{\mu}\> \in \spc H_{\mu}$, we obtain $A_{\mu}=
\<\psi_{\mu}|~\<A\>_{\grp G}~|\psi_{\mu}\>=\Tr_{\spc H_{\mu}} [A
~\<B_{\mu}\>_{\grp G}]$, where $B_{\mu}= |\psi_{\mu}\>\<\psi_{\mu}|
\otimes \openone_{\spc M_{\mu}}$. Now, since the group average
$\<B\>_{\grp G}$ is in the commutant of $\{U_g\}$, and since
$|\psi_{\mu}\> \in \spc H_{\mu}$, we have $\<B\>_{\grp G}= 1/d_{\mu}
\openone_{\spc H_{\mu}} \otimes \openone_{\spc M_{\mu}}$ for some
constant $d_{\mu}$. The constant $d_{\mu}$ is simply evaluated by
taking the expectation value of $\<B\>_{\grp G}$ with respect to a
normalized vector $|\phi_{\mu}\>|\alpha_{\mu}\>\in \spc H_{\mu}\otimes
\spc M_{\mu}$. \qed

\bigskip
{\bf Remark.} The \emph{formal dimension} of Eq. (\ref{FormDim}) is
not a property of the sole Hilbert space $\spc H_{\mu}$, but also of
the irreducible representation acting on it. Depending on the
particular irreducible representations, the same Hilbert space may
have different formal dimensions.
\subsection{An application: two indentical and two conjugated coherent states}
Here we give two examples about the estimation of coherent states of a
harmonic oscillator. Both cases involve the Abelian group of
displacements in the complex plane, with projective representation on
infinite dimensional Hilbert space $\spc H$ given by the
Weyl-Heisenberg group of unitary operators $\{D(\alpha)~=~e^{\alpha
a^{\dag} -\alpha^* a}~|~\alpha~\in~\Cmplx\}$, where $a^{\dag}$ and $a$
are creation and annihilation operators respectively. Since the group
is Abelian, it is obviously unimodular, a translation-invariant
measure being $\frac{\d^2 \alpha}{\pi}$ (here we put the constant
$\pi$ just for later convenience).

In the first example (two identical coherent states) we will consider
two identical copies of an unknown coherent state, while in the second
(conjugated coherent states) we will consider two coherent states with
the same displacement in position and opposite displacement in
momentum. Exploiting the method of maximum likelihood we will find in
both cases the optimal POVM for the estimation of the unknown
displacement. From the comparison between the sensitivities of the
optimal measurements in the two cases, a close analogy will emerge
with the well known example by Gisin and Popescu about quantum
information carried by parallel and anti-parallel
spins\cite{GisPop}. This analogy, already noticed in the study of the
optimal ``phase conjugation map'' by Cerf and Iblisdir\cite{CerIb},
will be analyzed here in detail from the general point of view of
group parameter estimation.
\subsubsection{Two identical coherent states}  
Here we consider two harmonic oscillators prepared in the same unknown
coherent state $|\alpha\>$, $\alpha \in \Cmplx$. In this case, the
family of signal states is
\begin{equation}
\mathcal{S}=\{~ |\alpha\>|\alpha\> \in \spc H^{\otimes 2}~|~ \alpha \in \Cmplx\}~,
\end{equation} 
and is obtained from the ground state
$|\Psi\>=|0\>|0\>$. by the action of the two-fold tensor representation
$\{D(\alpha)^{\otimes 2}~|~ \alpha \in \Cmplx\}$.  The Clebsch-Gordan decomposition of such a representation can be easily obtained by using the relation
\begin{equation}
D(\alpha)^{\otimes 2} = V^{\dag}~D(\sqrt{2}\alpha) \otimes
\openone~V~,
\end{equation}
where $V=\exp{\left[-\frac{\pi}{4} (a_1^{\dag}a_2-a_1
a_2^{\dag})\right]}$ ($a_1$ and $a_2$ denoting annihilation operators
for the first and the second oscillator respectively).  This relation
shows that---modulo a non-local change of basis in the Hilbert
space---the two-fold tensor representation is unitarily equivalent to
a direct sum where the irreducible representation $\{D(\sqrt{2}
\alpha)~|~\alpha \in \Cmplx\}$ appears with infinite
multiplicity. Such a representation is square-summable, and has the
formal dimension
\begin{equation}
d=\left( \int_{\Cmplx} \frac{\d^2 \alpha}{\pi}~ |\<0|~D(\sqrt{2}\alpha)~|0\>|^2 \right)^{-1}=2~,
\end{equation} 
given by Eq. (\ref{FormDim}). Moreover, according to
Eq. (\ref{DirectSum}), a possible decomposition of the tensor product
Hilbert space into irreducible subspaces is given by any set of the
form
\begin{equation}
\spc H_{n}= V^{\dag} ~\spc H \otimes |\phi_n\>~,   
\end{equation}
where $\{|\phi_n\>~|~n\in 0,1, \dots\}$ is an orthonormal basis for
$\spc H$. By taking the basis of eigenvectors of the number operator
$a^{\dag}a$, we immediately see that the input state
$|\Psi\>=|0\>|0\>$ completely lies in the irreducible subspace $\spc
H_0$. Denoting by $P_0=(V^{\dag} \openone \otimes |0\>\<0|) V$ the
projection onto $\spc H_0$, we have indeed $P_0|\Psi\>=|\Psi\>$.
Using Theorem \ref{TheoOptPOVM}, we have that for the state $|\Psi\>$
the optimal-likelihood covariant POVM must have $\Xi$ such that
$P_0\Xi P_0=|\eta\>\<\eta|$ with $|\eta\>=\sqrt{2}|0\>|0\>$, since
here $r_\mu=1$ (see Eq. (\ref{Schmidt})). Then, we have that any
covariant POVM with $P_0 ~\Xi~P_0=2 \left(|0\>\<0|\right)^{\otimes 2}$
is optimal for estimation of $\alpha$.  For example, we can take the
POVM
\begin{equation}
M(\alpha) = 2~D(\alpha)^{\otimes 2} ~\left(V^{\dag}~|\openone\>\!\>\<\!\<\openone|~V\right)~ D(\alpha)^{\dag \otimes 2 }~,
\end{equation}
where the unitary $V$ is defined as above, and $|\openone\>\!\>$ is
the vector $|\openone\>\!\>~=~\sum_{n=0}^{\infty}~|n\>|n\>$.  It can
be shown that this POVM corresponds to measuring the two commuting
observables corresponding to the position of the first oscillator and
the momentum of the second one. In this scheme, if the outcomes of the
two measurements are $q_1$ and $p_2$ respectively, we simply declare
that our estimate of the displacement is $\alpha= q_1+ip_2$.

A different POVM which is equally optimal is
\begin{equation}
\tilde M(\alpha)= 2~D(\alpha)^{\otimes 2} ~\left( V^{\dag}~|0\>\<0| \otimes \openone ~V\right)~D(\alpha)^{\dag \otimes 2 }~.
\end{equation}
In a quantum optical setup, this POVM corresponds to performing
firstly an heterodyne measurement on each oscillator, thus obtaining
two different estimates $\alpha_1$ and $\alpha_2$ for the
displacement, and then averaging them with equal weights. The final
estimate is $\alpha=\frac{(\alpha_1+\alpha_2)}2$.

Although the two POVM's are different and correspond to two different
experimental setups, they give rise to the same probability
distribution when applied to coherent states. It is indeed
straightforward to see that the probability density of estimating
$\hat \alpha$ when the true displacement is $\alpha$ is given in both
cases by the Gaussian
\begin{equation}\label{Gauss1}
p(\hat \alpha | \alpha)= 2~e^{-2|\hat \alpha -\alpha|^2}
\end{equation}
(normalized with respect to the invariant measure $\frac{\d^2
\alpha}{\pi}$). The value of the likelihood is $p(\alpha|\alpha)=2$,
according to our general formula (\ref{OptLik}).

\bigskip
{\bf Remark :} \emph{Improving the likelihood with squeezing.}\\ The
maximum likelihood can be improved using the doubly-squeezed state
\begin{equation}
|\Psi_x\>=V^\dag \sqrt{1-x^2} \sum_{n=0}^\infty x^n|n\>|n\>~,
\end{equation}
where without loss of generality we choose
$x>0$ ($x<1$ for normalization). Then, by applying  Theorem \ref{TheoOptPOVM}, it is immediate to show that $|\eta\>=\sqrt{2}V^\dag|I\>\!\>$
and to evaluate the  likelihood of the optimal POVM as
\begin{equation}
p(\alpha|\alpha)= 2~\frac{1+x}{1-x}~.
\end{equation}
Notice that for zero squeezing ($x=0$) we retrieve the case of two
identical coherent states, while for infinite squeezing ($x \to 1^-$)
the likelihood becomes infinite, according to the fact that the
displaces states $D(\alpha)^{\otimes 2}|\Psi_{x \to 1^-}\>$ become
orthogonal in the Dirac sense, allowing for an ideal estimation.
\subsubsection{Conjugated coherent states}

Now the family of signal states is
\begin{equation}
\mathcal S =\{|\alpha\>|\alpha^*\>~|~\alpha \in \Cmplx\}~,
\end{equation}
where complex conjugation is defined with respect to the basis
$\{|n\>~|~n=0, 1, \dots \}$. These states are generated from the input
state $|\Psi\>=|0\>|0\>$ by the action of the representation
$\{D(\alpha) \otimes D(\alpha^*)~|~ \alpha \in \Cmplx\}$.

Unfortunately, such a representation cannot be decomposed into a
discrete Clebsch-Gordan series, due to the fact that all the unitaries
in the representation can be simultaneously diagonalized on a
continuous set of non normalizable eigenvectors. In fact, for any vector of
the form $|D(\beta)\>\!\>=D(\beta) \otimes \openone |\openone\>\!\>$,
where $|\openone\>\!\>$ is the vector
$|\openone\>\!\>=\sum_{n}~|n\>|n\>$, we have
\begin{equation}
D(\alpha) \otimes D(\alpha^*)~|D(\beta)\>\!\>= e^{\alpha \beta^*-\alpha^* \beta}~|D(\beta)\>\!\>~.
\end{equation}
These vectors are orthogonal in the Dirac sense, namely
$\<\!\<D(\alpha)|D(\beta)\>\!\>= \pi \delta^2(\alpha -\beta)$.
Therefore, any such vector can be regarded the basis of a
one-dimensional irreducible subspace $\spc H_{\beta}$. The
multiplicity of any irreducible representation is one, and the Hilbert
space can be decomposed as a direct integral
\begin{equation}
\spc H \otimes \spc H= \int_{\Cmplx}^{\oplus}~ \frac{\d^2 \beta}{\pi}~ \spc H_{\beta}~.
\end{equation}
In the same way as in (\ref{CommutingOp}), an operator $O \in \Bnd
{\spc H \otimes \spc H}$ in the commutant of the representation can be
written as
\begin{equation}
O= \int_{\Cmplx}^{\oplus} \frac{\d^2 \beta}{\pi}~ \openone_{\beta}~o(\beta)~,
\end{equation}
where $\openone_{\beta}~=~|D(\beta)\>\!\>\<\!\<D(\beta)|$ is the identity in $\spc H_{\beta}$, and  $o(\beta)$ is some scalar function.

In this particular example it is easy to extend the results of
Sections \ref{OptMeasurements} and \ref{OptInputStates} to the case of
a direct integral of irreducible representations. In fact, using
functional calculus we can generalize the formula (\ref{AveOp}) for
the group average:
\begin{prop}
The average $\<A\>_{\Cmplx}$ of an operator over the representation
$\{D(\alpha)~\otimes~D(\alpha^*)~|~\alpha \in \Cmplx\}$ is
\begin{equation}
\<A\>_{\Cmplx}= \int_{\Cmplx}^{\oplus} \frac{\d^2 \beta}{\pi}~ \openone_{\beta} ~\Tr_{\spc H_{\beta}} [~A~]~,
\end{equation}
where $\Tr_{\spc H_{\beta}}[~A~]= \<\!\< D(\beta)|~A~|D(\beta)\>\!\>$.
\end{prop}
This expression for the group average is equivalent to that of
Eq.(\ref{AveOp}) modulo the obvious substitutions:
\begin{equation}
\left\{
\begin{array}{lll}
\bigoplus_{\mu \in \set S} &\to& \int^{\oplus}_{\Cmplx} \frac{\d^2 \beta}{\pi}\\
&\\
d_{\mu} &\to& d_{\beta}=1 \qquad \forall \beta \in \Cmplx
\end{array}
\right.
\end{equation} 
The optimal POVM is obtained by Theorem \ref{TheoOptPOVM} by making
these substitutions. We just need to decompose the input state on the
irreducible subspaces, i.e.
\begin{equation}
|0\>|0\>= \int^{\oplus}_{\Cmplx} \frac{\d^2 \beta}{\pi}~
e^{-|\beta|^2/2}~|D(\beta)\>\!\>~,
\end{equation}
and then take the optimal POVM given by the operator
$\Xi=|\eta\>\<\eta|$ in Eqs. (\ref{OptXi}) and (\ref{OptEta}), which
in the present case becomes
\begin{equation}
|\eta\>= \int_{\Cmplx}^{\oplus} \frac{\d^2 \beta}{\pi}~ |D(\beta)\>\!\>
~.
\end{equation}
Notice that, since the input state $|0\>|0\>$ has nonzero components
in all the irreducible subspaces, the optimal covariant POVM is now
unique. Using such optimal POVM,
\begin{equation}
M(\alpha)= D(\alpha) \otimes D(\alpha^*) ~|\eta\>\<\eta|~
D(\alpha)^{\dag} \otimes D(\alpha^*)^{\dag}~,
\end{equation} 
the probability density of estimating $\hat \alpha$ when the true
displacement is $\alpha$ can be calculated to be the Gaussian
\begin{equation}\label{Gauss2}
p(\hat \alpha|\alpha)= 4~ e^{-4 |\hat \alpha-\alpha|^2}~
\end{equation}
(normalized with respect to the invariant measure $\frac{\d^2 \alpha}{\pi}$).

Notice that the value of the likelihood $p(\alpha|\alpha)=4$ could
also be calculated directly using the formula (\ref{OptLik}), which
now reads
\begin{equation}
p(\alpha|\alpha)= \left( \int_{\Cmplx} \frac{\d^2 \beta}{\pi}~
e^{-|\beta|^2/2} \right)^2=4.
\end{equation}
Comparing the optimal distribution (\ref{Gauss2}) for two conjugated
coherent states with the corresponding one for two identical coherent
states (\ref{Gauss1}) we can observe that the variance has been
reduced by one half, while the likelihood has become twice. It is
interesting to note the remarkable analogy between this example in
continuous variables and the example by Gisin and Popescu
\cite{GisPop} about the quantum information encoded into a pair of
parallel and anti-parallel spins.  In fact, in the case of spins the
authors stressed that, quite counter-intuitively, while two classical
arrows pointing in opposite direction carry the same information, in a
quantum mechanical setup two anti-parallel spins carry more
information than two parallel ones. In the same way, in the continuous
variables context, while classically two conjugated points $\alpha$
and $\alpha^*$ in the phase space carry the same information (such
information being the couple of real numbers $(x,p)$~), quantum
mechanically two conjugated coherent states carry more information
than two identical ones. The analogy is even closer, since for
spin-$\frac{1}{2}$ particles the ``spin-flip'' operation is unitarily
equivalent to the complex conjugation, whence we can regard also the
example of spins as a comparison between pairs of identical states and
pairs of conjugated states.

It is important to stress that the group theoretical analysis and the
maximum likelihood approach provide in both cases also a clear
explanation of the mechanism generating the asymmetry between pairs of
identical and conjugated states. In fact, the whole orbit of a pair of
identical coherent states (or spins states) lies just in \emph{one}
irreducible subspace of the Hilbert space, while the orbit of a pair
of conjugated coherent states (spin states) covers \emph{all}
irreducible subspaces. According to formula (\ref{OptLik}), the
likelihood in the case of conjugated states is higher than the
likelihood for identical states both in the case of coherent and spin
states, corresponding to an enhancement of the probability of
successful discrimination.

\section*{Acknowledgments}This work has been supported by the FET European
Networks on Quantum Information and Communication Contract
IST-2000-29681:ATESIT, by MIUR 2003-Cofinanziamento, and by INFM
PRA-2002-CLON.

\end{document}